\documentclass[10pt,final,superscriptaddress,amsmath,amssymb,aps,prd,twocolumn,showpacs,nofootinbib]{revtex4-2}

\usepackage{natbib}
\usepackage{amsmath}
\usepackage{mathrsfs}
\usepackage{amsthm}
\usepackage{physics}
\usepackage{float}
\usepackage{hyperref}
\usepackage{makecell}
\usepackage{graphicx}
\usepackage{dcolumn}  
\usepackage{bm}  
\usepackage{xcolor}

\definecolor{citecolor}{rgb}{1.0000,    0.2078,    0.3686}
\newcommand{\paramspace}{ \mathcal{P}_{\mathrm{KN}} }
\newcommand{\iu}{ \, {i}}

\begin{document}

\title{Exceptional points and hysteresis in Kerr-Newman quasi-normal modes}

\author{Mario Misas Arcos}
\thanks{Contact author: mmisas@us.es} %
\affiliation{Departamento de Electrónica y Electromagnetismo, Universidad de Sevilla, Facultad de Física, 41012 Sevilla, Spain}

\author{Alberto T. P\'erez}%
\thanks{Contact author: alberto@us.es}
\affiliation{Departamento de Electrónica y Electromagnetismo, Universidad de Sevilla, Facultad de Física, 41012 Sevilla, Spain}

\author{Carlos F. Sopuerta}
\thanks{Contact author: carlos.f.sopuerta@csic.es}
\affiliation{Institut de Ciències de l’Espai (ICE, CSIC), Campus UAB,
Carrer de Can Magrans s/n, 08193 Cerdanyola del Vallès, Spain}
\affiliation{Institut d'Estudis Espacials de Catalunya (IEEC), Edifici RDIT, C/ Esteve~Terradas, 1, desp.~212, Castelldefels~08860, Spain}

\date{\today}

\begin{abstract}    
Exceptional points (EPs) and eigenstate hysteresis, previously found in Kerr only with an auxiliary scalar field, arise intrinsically in the quasinormal-mode spectrum of Kerr-Newman black holes. 
Using a hyperboloidal formulation of the coupled gravito-electromagnetic Teukolsky equations and a Chebyshev pseudospectral method, we compute the $\ell=m=2$ spectrum and locate four EPs at up to $99.974\%$ of extremality, one from a new repulsion. 
The gaps close as a square root branch point, and quasistatic loops around one or more EPs exchange the coalescing modes, confirming hysteresis.
\end{abstract}


\maketitle


\noindent\textit{Introduction.}~A black hole (BH), as described by general relativity (GR), is a region of spacetime from which nothing can ever escape to reach very distant observers. As such, BHs are purely geometry objects characterized by only three parameters as shown by the no-hair theorems~\cite{Israel:1967wq,Carter:1971zc,Robinson:1975bv,Mazur:1982db} (see also~\cite{Chrusciel:2012jk})\footnote{Throughout this work, we use natural units in which $c = G = 1\,$.}: mass $M$, spin $J/M=a$, and electric charge $Q$; with $0\leq a,Q\leq M$ and $a^2 + Q^2 < M^2$. A perturbed isolated BH relaxes through a discrete spectrum of damped sinusoids, its quasi-normal modes (QNMs), with frequencies fixed by $M$, $a$, and $Q$ alone~\cite{Regge:1957td,Vishveshwara:1970zz,Leaver:1985ax,Kokkotas:1999bd,Berti:2009kk}.
QNMs dominate the post-merger ringdown, and they have been already identified in LIGO-Virgo-KAGRA events as GW150914~\cite{LIGOScientific:2016aoc,LIGOScientific:2016lio} and GW250114, the loudest event observed to date, yielding the most stringent single-event test of the Kerr hypothesis so far~\cite{LIGOScientific:2025wao}. Because two or more resolved modes overdetermine the remnant's parameters, any inconsistency among them signals a departure from GR or from the Kerr solution itself: this is the program of BH spectroscopy~\cite{Detweiler:1980gk,Dreyer:2003bv,Berti:2005ys}.

The Kerr-Newman (KN) metric~\cite{Newman:1965my,Adamo:2014baa}, the unique stationary electrovacuum BH solution, reduces to Kerr~\cite{Kerr:1963ud} at $Q=0$ with extremality reached at $M^2=a^2+Q^2$~\cite{Chandrasekhar:1983bbh}. The event horizon radius is given by $r_+=M+\sqrt{M^2-a^2-Q^2}$. 
Like Kerr, KN is an algebraically-special (Petrov type D) spacetime. Astrophysical BHs are expected to be essentially neutral~\cite{Gibbons:1975kk}, yet KN remains the natural arena for intrinsically coupled gravito-electromagnetic perturbations, and hence for the multi-field spectral phenomena we report in this letter.

Linear perturbations of spinning BHs are most naturally studied in the Newman-Penrose formalism~\cite{Newman:1961qr}, where Teukolsky showed that the curvature and Maxwell scalars $(\Psi_0,\Psi_4,\Phi_0,\Phi_2)$ obey decoupled, separable master equations on a  Kerr background~\cite{Teukolsky:1973ha} (see also~\cite{Chandrasekhar:1983bbh}). For a KN BH background,  while the Klein-Gordon and Dirac equations are separable~\cite{Berti:2009kk}, it has not until very recently that the gravito-electromagnetic perturbations have been separated although not fully decoupled~\cite{Hintz:2026jiw} (see also~\cite{Destounis:2026ljd,Chakraborty:2026kyf,Motohashi:2026wac} for applications of this breakthrough).  The nonzero background electromagnetic field couples the spin-2 and spin-1 sectors directly in the perturbation equations and closed-form separation of this coupled system was not achieved before (see~\cite{Chandrasekhar:1983bbh}). Previously, some progress was made~\cite{Berti:2005eb} through the non-self-consistent Dudley-Finley decoupling~\cite{Dudley:1977zz}, perturbative expansions in $a$ or $Q$~\cite{Mark:2014aja}, or, most recently, direct numerical solution of the genuinely coupled two-field boundary-value problem~\cite{Dias:2021yju,Dias:2022oqm}, whose well-posedness has been established rigorously in restricted parameter regimes~\cite{Giorgi:2023fqi,Giorgi:2025ylc}.

The KN QNM spectrum has been studied since the work of Kokkotas~\cite{Kokkotas:1993ef} and Berti and Kokkotas continued-fraction solution of the separable scalar sector, which also validated the Dudley-Finley approximation for the coupled gravito-electromagnetic fields at moderate charge~\cite{Berti:2005eb} (see~\cite{Dias:2015wqa,Destounis:2026ljd} for recent work). There are a number of works containing analytic, WKB, and special-limit studies~\cite{Jing:2005pk,Jing:2008zzb,Kokkotas:2010zd,Couch:2012zz,Konoplya:2013rxa,Hod:2014uqa,Hod:2015xlh,Zhao:2015pqa}. Near-extremal KN was addressed analytically in~\cite{Hod:2008se}, and the fully coupled fields perturbatively at weak charge in~\cite{Mark:2014aja}. A decisive advance came with the direct numerical solution of the coupled, non-separable gravito-electromagnetic system~\cite{Dias:2021yju,Dias:2022oqm}, establishing the mode stability of KN BHs and its characteristic eigenvalue repulsion between photon-sphere and near-horizon mode families, subsequently confirmed for scalar perturbations~\cite{Davey:2023fin}, and placed on a rigorous mathematical footing~\cite{Giorgi:2023fqi,Giorgi:2025ylc}. Recently, this spectrum has been revisited with physics-informed neural solvers~\cite{Pombo:2026xsj}, exceptional lines in scalar perturbations~\cite{Cavalcante:2026vgr}, a critical reassessment of the Dudley-Finley approximation~\cite{Saha:2025nsg}, and a  correspondence with Seiberg-Witten theory~\cite{Silva:2025khf}.

Exceptional points (EPs) are singularities in the parameter space of non-Hermitian operators at which two or more eigenvalues and their eigenvectors coalesce~\cite{Kato:1995cw}, so that the operator does not have a complete set of eigenvectors~\cite{Heiss:2012,Miri:2019}. 
In their neighborhood, the eigenvalues behave as square-root branch points, so that a 
closed loop in parameter space encircling an EP exchanges the coalescing states, which do not return to their initial configuration, a form of eigenstate hysteresis~\cite{Heiss:2012,Doppler:2016}. EPs are relevant for
physical phenomena in systems such as microwave resonators~\cite{Dembowski:2001prl} and open quantum systems~\cite{Choi:2021prl}.

EPs in the BH QNM spectrum have so far only been found in cases where an auxiliary massive test scalar field has been introduced in a Kerr background. The mass of this scalar field constitutes the second parameter required for a genuine two-parameter non-Hermitian spectral problem~\cite{Cavalcante:2024swt}. Whether this phenomenology can instead arise intrinsically, from the BH own multi-field perturbation structure, and how the coupled KN-Teukolsky system should be formulated in a horizon-to-infinity, boundary-condition-free setting, have remained open. We address both questions in this letter.

Since EPs of a generic non-Hermitian problem have real co-dimension two, they appear as isolated points in a two-parameter space. One can locate EPs in KN BHs by varying the coordinates of the space spanned the dimensionless spin and charge,
$\mathcal{P}_{\mathrm{KN}}=\{(a/M,\,Q/M)~\mathrm{with}~(a/M)^2+(Q/M)^2\le 1\}$, which makes them an ideal area to explore EPs in connection with QNMs. Following~\cite{Dias:2021yju,Dias:2022oqm}, we can adopt alternative coordinates normalized with respect to the horizon radius: $\hat{a} = a/r_+$ and $\hat{Q} = Q/r_+$. In these coordinates, the extremal limit is achieved when: $\hat{a}_{\text{ext}} = (1 - \hat{Q}^2)^{1/2}$. We can explore the space $\mathcal{P}_{\mathrm{KN}}$, scanning $\hat{Q} \in [0,1]$ and following curves of constant ratio
\begin{equation}
k \equiv \frac{\hat{a}}{\hat{a}_{\text{ext}}} = \frac{\hat{a}}{\sqrt{1 - \hat{Q}^2}} \,.
\label{eq-def-k-ratio}    
\end{equation}

In this letter  we focus on \textit{eigenstate hysteresis}: the non-trivial interchange of coalescing eigenstates when the system is subject to parametric changes following a quasi-static trajectory that encircles an EP.

~

\noindent\textit{Problem statement.}
In this letter, we explore the KN sub-extremal parameter space $\mathcal{P}_{\mathrm{KN}}$ and accurately map four EPs in the QNM spectrum
and restrict our analysis to the dominant of the quadrupolar modes $\ell = m = 2$.
We report and study the presence of eigenstate hysteresis around all four EPs found, allowing us to draw unique connections between BH QNM spectroscopy and general non-Hermitian spectral topology. 

The QNM computations are enabled by building upon the \textit{hyperboloidal foliation} framework developed in~\cite{PanossoMacedo:2024nkw} for this purpose (see also~\cite{zengi_og,zengi_framework,panossokerr,kerrqnm2d}).
The present work constitutes the first reported application to the study of the KN QNM spectrum, rendering the location of EPs in the KN QNM spectrum accessible.
The key point is that on constant-$t$ slices (being $t$ the Boyer-Lindquist time coordinate~\cite{Boyer:1966qh}), QNM eigenfunctions diverge exponentially toward the horizon and spatial infinity, so outgoing boundary conditions must be imposed on a truncated domain. Hyperboloidal slices instead cross the event horizon and reach future null infinity $\mathscr{I}^+$~\cite{zengi_og,zengi_framework}. QNM eigenfunctions are then regular, boundary conditions reduce to regularity, and the radial domain becomes finite~\cite{panossokerr,kerrqnm2d}. 
We extend the Kerr construction of Ref.~\cite{kerrqnm2d} to KN. Let $(t,r,\theta,\varphi)$ be KN Boyer-Lindquist coordinates, with $\Delta=r^2-2Mr+a^2+Q^2$ and $\Sigma_0=r^2+a^2$.
We can introduce ingoing KN coordinates $(v,r,\theta,\phi)$, which are regular at the future horizon:  
$v=t+r^*(r)$ and $\phi=\varphi+k(r)$, where $\mathrm{d}r^*/\mathrm{d}r=\Sigma_0/\Delta$ and $\mathrm{d}k/\mathrm{d}r=a/\Delta$. We then compactify and bend the slices up to $\mathscr{I}^+$ by introducing $(\tau,\sigma,\theta,\phi)$, with
\begin{equation}
v=\lambda [\tau-h(\sigma)], ~
r=\frac{\lambda}{\sigma},\quad 
h(\sigma)=2\left[\frac{2M}{\lambda}\ln\sigma-\frac{1}{\sigma}\right],
\label{eq:chart}
\end{equation}
where $\tau$ is the hyperboloidal time, $\sigma\in(0,1]$ is the compactified radial coordinate, $h$ is the height function, and $\lambda=r_+$ is a length scale chosen so that $\sigma=1$ corresponds to the event horizon and $\sigma=0$ to $\mathscr{I}^+$.
The choice of $h(\sigma)$ and of $r(\sigma)$ encodes the gauge freedom of the construction, and Eq.~\eqref{eq:chart} is the minimal choice that fixes the horizon at $\sigma=1$.
In these coordinates the radial and angular operators of the coupled Teukolsky system become first-order in $\partial_\sigma$ and $\partial_\tau$ with coefficients regular on the whole domain. See~\cite{Dias:2015wqa} for their explicit form.

For the numerical implementation we adopt the pseudospectral collocation method, frequently used for BH perturbation theory and QNM computations~\cite{Canizares:2009ay,Canizares:2010yx,panossokerr,Dias:2015wqa} (see also~\cite{Boyd:2013}).  We use a multiple Chebyshev-Lobatto grid, $\{(X_i,Y_j)\in[-1,1]\times [-1,1]\}$ (i.e., containing the boundary points), associated with the hyperboloidal foliation coordinates $(\sigma,\theta)$, being the mapping given by: $\sigma_i = (X_i + 1)/2$ and $\cos{\theta_j} = Y_j$. Once discretized, the resulting system can be written as a block-matrix polynomial eigenvalue problem that we solve using the two-level orthogonal Arnoldi method implemented in the SLEPc library~\cite{Hernandez:2005:SSF,Campos:2016:PKS}  (see~\cite{Hernandez:2007:PAE,Roman:2023:ISR,Misas:2026fp} for further details). See the companion paper~\cite{Misas:2026fp} for the details on the implementation of the computations shown in this letter.

~

\noindent\textit{EPs and eigenvalue repulsion}.
Work on EPs in BH QNM spectra has recently flourished (see~\cite{Cavalcante:2025abr,Yang:2025dbn,Motohashi:2024fwt}). In~\cite{Yang:2025dbn}, it is shown that resonant frequencies near an EP induce a linearly-growing prefactor in the time-domain ringdown signal which, if unaccounted for in waveform models, may introduce a bias in the extraction of the QNM. In the KN case~\cite{Dias:2021yju,Dias:2022oqm}, we have two tunable parameters, the phenomenon of  eigenvalue repulsion (avoided crossings) has recently been found (see~\cite{Rossi:2026als} for the Kerr-de Sitter case). The phenomenon of eigenvalue repulsion appears in quantum physics, typically known as the Wigner-Teller effect~\cite{landau1977quantum}. It is also found in a broader context in bifurcation theory, where mode paths that break the same symmetry avoid crossing (for instance, in stability problems in hydrodynamics \cite{Vandevooren:1989had,atp:2014}).

\begin{figure}[]
{\centering
\includegraphics[width=0.95\linewidth]{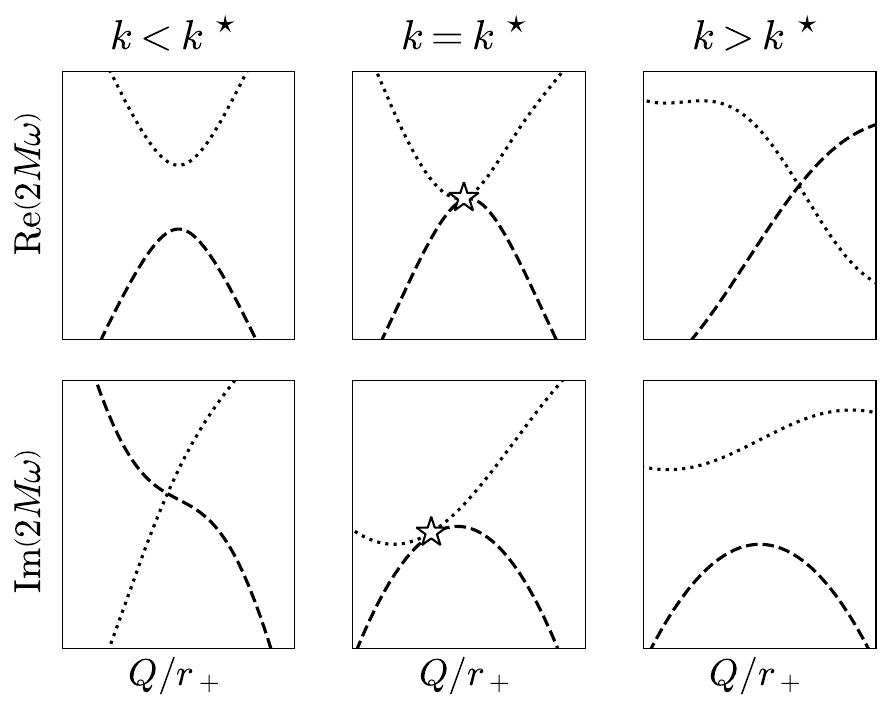}}
\caption{ 
Behavior of eigenvalue branches depending on the parameters $k$ and $\hat{Q}$ near an EP [See Eq.~\eqref{eq-def-k-ratio}]. Here, $k^\star$ denotes the value of $k$ at the location of the EP.
}
\label{fig:qualit}
\end{figure}

We schematically show in Figure~\ref{fig:qualit} the evolution of eigenvalues near an EP. 
The real and imaginary parts of two modes are plotted as a function of $\hat{Q}$ while keeping $k$ constant.
Below the critical value, the imaginary parts intersect, while the real parts exhibit eigenvalue repulsion (avoid crossing). At an EP, both the real and imaginary parts intersect. Beyond the EP, they 
exchange roles. 
This behavior is associated with the solutions of a quadratic equation in the complex plane. The EP corresponds to a branching point with two associated Riemann sheets. This implies the possibility of hysteresis. If we describe a closed path in the parameter space, whether we move from one sheet to another or remain in the same sheet depends on whether we enclose or not the branching point \cite{Heiss:2012}.

\begin{figure}[]
\centering
\includegraphics[width=0.95\linewidth]{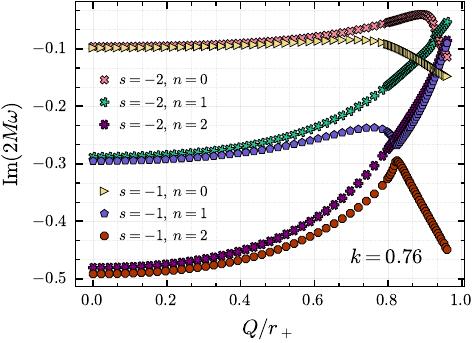}
\caption{Parametric evolution of six QNM branches along a $k = 0.76$ constant curve. Gravitational modes (in the Kerr limit) are marked with asterisks, while polygonal markers are used for electromagnetic modes. Avoided crossing is clearly visible between the branches $(s,n) = (-1,1)$ and $(-1,2)\,$.}
\label{fig:branches}
\end{figure}

We have identified, within the KN QNM spectrum, the precise location of the three instances of EPs responsible for the avoided crossing reported in~\cite{Dias:2022oqm} (first three rows of Table~\ref{tab:expoints}). The fourth one is a new case of repulsion, happening between electromagnetic-led branches (see Fig.~\ref{fig:branches}).

\normalsize
\begin{table}[]
\centering
\renewcommand{\arraystretch}{2}
\setlength{\tabcolsep}{5pt}
\begin{tabular}{|c|c|c|c|c|}
\cline{2-5}
\multicolumn{1}{c|}{} & $\operatorname{Re} \left( 2M \omega \right)$ & $\operatorname{Im} \left( 2M \omega \right)$ & $a^\star/M$     & $Q^\star/M$  \\ 
\hline
EP$^{}_0$ &  \makecell{1.12869 \\ $\pm$ 0.00093} & \makecell{$-0.056626$ \\ $\pm$ 0.000040} & \makecell{0.304065 \\ $\pm$ 0.000029} & \makecell{0.952379\\ $\pm$ 0.000020}  \\ 
\hline
EP$^{}_1$ & \makecell{0.9091 \\ $\pm$ 0.0024}   & \makecell{$-0.3704$ \\ $\pm$ 0.0050}     &\makecell{ 0.220127 \\ $\pm$ 0.000076} &  \makecell{0.972975 \\ $\pm$ 0.000076}  \\ 
\hline
EP$^{}_2$ & \makecell{0.8984 \\ $\pm$ 0.0058}   & \makecell{$-0.3833$ \\ $\pm$ 0.0038}     & \makecell{0.20930 \\$\pm$ 0.00026} & \makecell{0.97145 \\$\pm$ 0.00041} \\ 
\hline
EP$^{}_e$ & \makecell{1.5863 \\ $\pm$ 0.0025}   & \makecell{$-0.2900$ \\ $\pm$ 0.0018}     & \makecell{0.461164 \\$\pm$ 0.000056} & \makecell{0.884039 \\$\pm$ 0.000055} \\ 
\hline
\end{tabular}
\caption{Frequencies and coordinates of four EPs. 
}
\label{tab:expoints}
\end{table}

We track the parametric evolution across $\paramspace$ of six different eigenstates within the KN QNM spectrum: the fundamental mode and the first and second overtones of both the gravitational ($s = -2$) and electromagnetic ($s = -1$) $\ell = m = 2$ states. 
The appearance of an EP in $\paramspace$ is signaled by the emergence of the behavior shown in Fig.~\ref{fig:qualit}. 
Starting from $k=0$, we construct \textit{branches} of QNMs that evolve continuously, along curves of constant $k$, from the states with $\hat{Q} = 0$ (see Fig.~\ref{fig:branches}). Since $\hat{Q} = 0$ implies $Q = 0$ for any value of $k$, the first mode in each branch must agree with the well-documented~\cite{Berti:2005ys} Kerr QNM spectrum for the corresponding value of $a$.

We denote EPs in $\paramspace$ as: $\mathrm{EP}_i \equiv (a^\star_i, Q^\star_i)$ (see Table~\ref{tab:expoints}; with $i=0,1,2,e$) or alternatively with horizon-normalized coordinates $(k^\star,\hat{Q}^\star)$. Then, we define the \textit{repulsion gap widths} $\Delta \mathfrak{I}$ and $\Delta\mathfrak{R}$ as the difference between the imaginary or real parts of the closest QNM frequencies within two branches, when said components (imaginary or real) exhibit avoided crossing. At an EP, with coordinates $(a^\star, Q^\star) \in \paramspace$, the repulsion-crossing behavior must satisfy: 
\begin{equation}
\lim_{(a,Q) \to (a^\star, Q^\star)} \Delta\mathfrak{I} =  \lim_{(a,Q) \to (a^\star, Q^\star)} \Delta\mathfrak{R} = 0 \,.
\label{eq:expoint}
\end{equation}
A total of $90$ $k$-constant parameter-sweeps have been performed, with $k \in \{0, 0.01, \dots 0.88 , 0.89\}$, where the evolution of the six specified eigenstates was tracked through a series of $64$ evenly-spaced values within the interval $\hat{Q} \in [0, 0.96]$.
We have identified four distinct regions of $\paramspace$ where EPs may be located, in light of Wigner-Teller effect being exhibited within four different intervals of $k$: 
(i) The repulsion extending within the broadest range is prominent within $k \in [0.81, \, 0.86]$, between branches $(-2,\,0)$ and $(-2,\,1)$.
We call $\mathrm{EP}_0$ the exceptional mode associated with the EP responsible for this behavior.
(ii) Branches $(-1,\,1)$ and $(-1,\,2)$ exhibit repulsion within the range $k \in [0.75, \, 0.77]$ (visible in Fig.~\ref{fig:branches}). Due to the \textit{electromagnetic} nature of these branches, we call the QNM associated with this EP, $\mathrm{EP}_e$.
(iii) For $k \in [0.49, 0.51]$, the $(-2,2)$ branch shows eigenvalue repulsion with a non-tracked family of modes, named \textit{near-horizon} modes in~\cite{Dias:2021yju}. This is a family of eigenstates whose wave function is mostly localized in the vicinity of the event horizon, and which are only found at relatively high levels of extremality. We call the QNM associated with this EP candidate $\mathrm{EP}_1$.
(iv) Branches $(-2,\,0)$ and $(-2,\,2)$ exhibit repulsion within the range $k \in [0.38, \, 0.41]$. We call the QNM associated with this EP candidate $\mathrm{EP}_2$.
Three of the four repulsions identified ($\mathrm{EP}_0$-$\mathrm{EP}_2$) were also reported in~\cite{Dias:2021yju}, the exception being the one hinting the existence of EP$^{}_e$.

\begin{figure}[]
\centering
\includegraphics[width=0.95\linewidth]{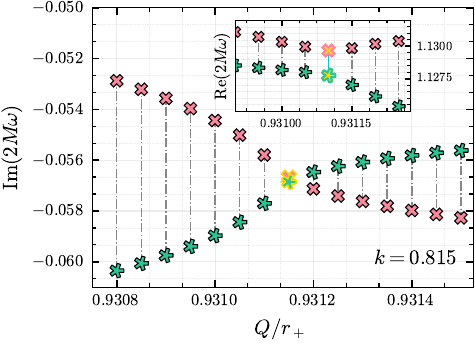}
\caption{High-resolution sampling of the branches $(s,n) = (-2,0)$ and $(-2,1)$ in the region $(k, \hat{Q}) \in \{0.815\} \times [0.9305, 0.9315]$. 
Unlike in Fig.~\ref{fig:branches}, the avoided crossing occurs in the real part; the cyan line in the inset shows $\Delta\mathfrak{R}$.
}
\label{fig:815}
\end{figure}

\begin{figure}[]
\centering
\includegraphics[width=0.95\linewidth]{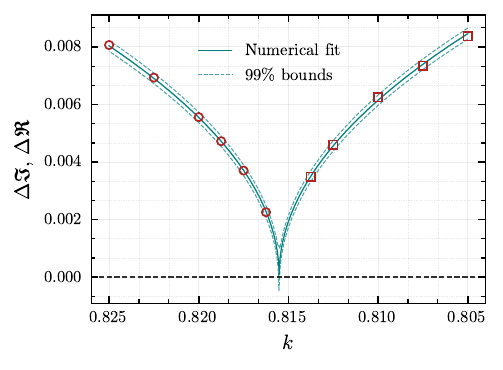}
\caption{Values of the repulsion gap widths (circles correspond to $\Delta \mathfrak{I}$ and squares to $\Delta \mathfrak{R}$), obtained by sampling in the vicinity of $k^\star$. 
The exceptional value $k^\star$ is where the fit crosses the dashed line. The fit is consistent with the square-root singularity expected at an EP~\cite{Heiss:2012}.
}  
\label{fig:kfit}
\end{figure}

\begin{figure}[]
\centering
\includegraphics[width=0.95\linewidth]{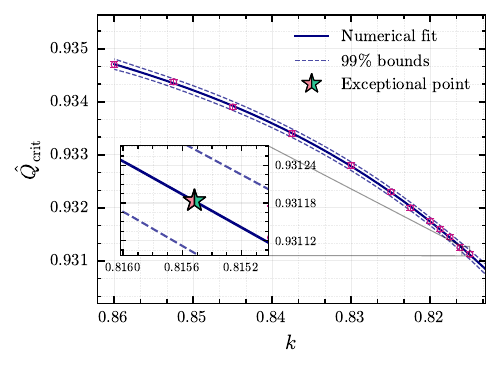}
\caption{Values of critical relative charge retrieved by sampling for $k > k^\star$. The EP color scheme is chosen regarding the eigenvalue branches on which it induces coalescence.} 
\label{fig:Qfit}
\end{figure}

To locate each EP precisely, we sampled the parameter space at higher resolution near the avoided crossings and estimated the gaps $\Delta\mathfrak{I}$ and $\Delta\mathfrak{R}$ (see Fig.~\ref{fig:815}), together with $\hat{Q}_{\mathrm{crit}}$, the value of $\hat{Q}$ at which the branches are closest to coalescence, for each $k$. Fitting the gaps to a square-root profile, as expected near an EP (see Fig.~\ref{fig:kfit}), gives $k^\star$ as the zero of the fit under the conditions in Eq.~\ref{eq:expoint}. A cubic fit of $\hat{Q}_{\mathrm{crit}}(k)$ (see Fig.~\ref{fig:Qfit}) then gives $\hat{Q}^\star\equiv\hat{Q}_{\mathrm{crit}}(k^\star)$.

~

\noindent\textit{Eigenstate hysteresis.}
Hysteresis of a QNM state is defined as the path-dependence of its evolution when following quasi-static paths in $\paramspace$ with equal endpoints. The presence of EPs induces a specific type of hysteresis, which is the non-trivial exchange of coalescing modes: tracking their evolution along a closed path in the parameter space will result in the modes being exchanged if and only if the path encircles the EP, at which they coalesce, an odd number of times. 
This phenomenon is connected to the underlying topology of the parameter space~\cite{Heiss:2012}. 
More specifically, to the fact that EPs  are topologically equivalent to square-root singularities of the system QNMs.

\begin{figure}[]
\centering
\includegraphics[width=0.95\linewidth]{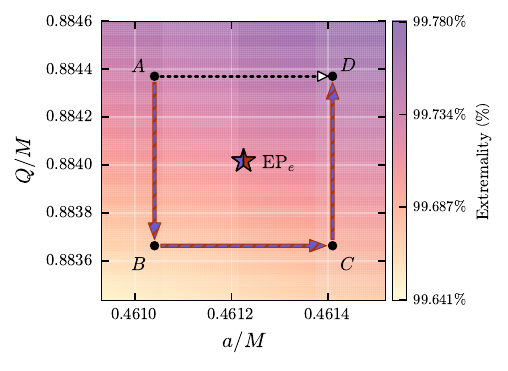}
\caption{Closed, quasi-static path around $\mathrm{EP}_e$ that induces hysteresis on the states $(s,n)=(-1,1)$ and $(-1,2)$ coalescing at $\mathrm{EP}_e$. The dashed line from $A$ to $D$ indicates the variation by following $\Gamma_{AD}$ in contrast to $\Gamma_{ABCD}$.}
\label{fig:paths}
\end{figure}

Observing eigenstate hysteresis serves a dual purpose: it rigorously confirms the existence and location of the presumed EPs, and it allows to map the non-trivial topology of the QNM parameter space, $\paramspace \times \{2 M\omega\}$, in the vicinity of an EP. 
Let us begin by analyzing hysteresis around $\mathrm{EP}_e$, with coordinates $(a^\star_e, Q^\star_e) = (0.461164, 0.884039) M$. We construct a closed contour defined by four points in $\paramspace$ (see Fig.~\ref{fig:paths}): 
$A = (0.9996 a_e^\star, 1.0004 Q_e^\star)$, 
$B = (0.9996 a_e^\star, 0.9996 Q_e^\star)$,
$C = (1.0004 a_e^\star, 1.0004 Q_e^\star)$, 
$D = (1.0004 a_e^\star, 0.9996 Q_e^\star)$.
Then, we track the evolution by varying the parameters $a/M$, $Q/M$ along two contours with identical endpoints: $\Gamma_{AD} \equiv A \to D$ and $\Gamma_{ABCD} \equiv A \to B \to C \to D$.
The choice of points ensures that the union $\Gamma_{ABCD} \cup \Gamma_{AD}$ encloses $\mathrm{EP}_e$ as represented in Fig.~\ref{fig:paths}. 
We use the subscripts $\alpha$ and $\beta$ to denote the $(s,n)=(-1,1)$ and $(-1,2)$ modes, respectively.

\begin{figure}[]
\centering
\includegraphics[width=0.95\linewidth]{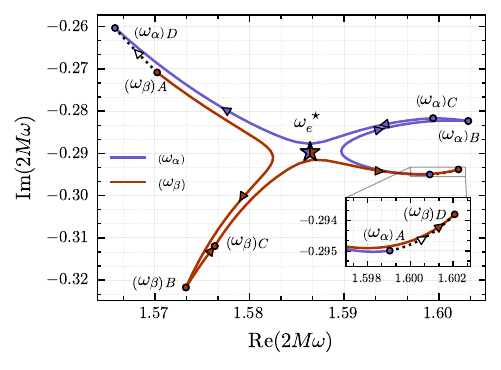}
\caption{Evolution of the QNM frequencies tracked along $\Gamma_{ABCD}$ (solid lines) and $\Gamma_{AD}$ (dashed lines). Here, we use $\omega_\alpha$ to denote frequencies corresponding to the $(-1,1)$ QNM branch, and $\omega_\beta$ to denote those corresponding to the $(-1,2)$ branch. Non-trivial exchange of states is present, confirming hysteresis has been successfully induced.}
\label{fig:paths_complex}
\end{figure}

At the starting point $A$, we find: $(\omega_\alpha)_{A} = 1.59904 - 0.29499 \iu$ and $(\omega_\beta)_{A} = 1.57025-0.27084 \iu$.
Following the path $\Gamma_{ABCD}$ yields the transformation: 
$(\omega_{\alpha, \beta})_{A} \to \cdots \to (\omega_{\alpha, \beta})_{C} \to (\omega_{\alpha, \beta})_{D}$ in such a way that 
$(\omega_\alpha)_{C} \mapsto (\omega_\alpha)_{D} = 1.56579 - 0.26032 \iu\,$,
$(\omega_\beta)_{C}  \mapsto (\omega_\beta)_{D} =  1.60207 - 0.29380 \iu\,$.
On the other hand, evolving along $\Gamma_{AD}$ produces the transformation:
$(\omega_{\alpha, \beta})_{A} \to (\omega_{\alpha, \beta})_{D}$, such that
$(\omega_\alpha)_{A} \mapsto (\omega_\alpha)_{D} = 1.60207 -0.29380 \iu$ and
$(\omega_\beta)_{A}  \mapsto (\omega_\beta)_{D} =  1.56579 - 0.26032 \iu$.
This confirms the path dependence of the state evolution in the form of non-trivial exchanges, as illustrated in Fig.~\ref{fig:paths_complex}. We have also observed this expected behavior for the other EPs in Table~\ref{tab:expoints} ($\mathrm{EP}_0$-$\mathrm{EP}_2$)
using analogous contours, but for brevity we do not give all details here (see~\cite{Misas:2026fp}).

\begin{figure}[]
\centering
\includegraphics[width=0.95\linewidth]{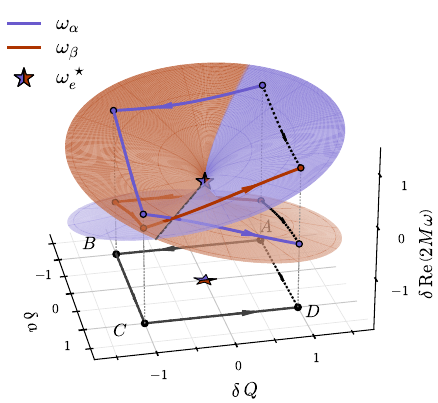}
\caption{Qualitative representation of the (real-part-projection of) $\paramspace \times \{ 2M \omega \}$ in the vicinity of EP$^{}_e$. The exchange of states is shown to correspond to a change of Riemann sheet (indicated by the color-mapping), allowed by the self-intersection of the Riemann surface between $C$ and $D$, where the real parts of $\omega_\alpha$ and $\omega_\beta$ undergo crossing.}
\label{fig:hysteresis}
\end{figure}

The topology of $\paramspace \times \{2M \omega\}$ near each EP is shown to be equivalent to that of a pair of Riemann sheets connected at the EP (typical of a square-root branch point). As anticipated, exchange of states corresponds to switching from one of the sheets to the other as shown in Fig.~\ref{fig:hysteresis}.

\begin{figure}[]
\centering
\includegraphics[width=0.95\linewidth]{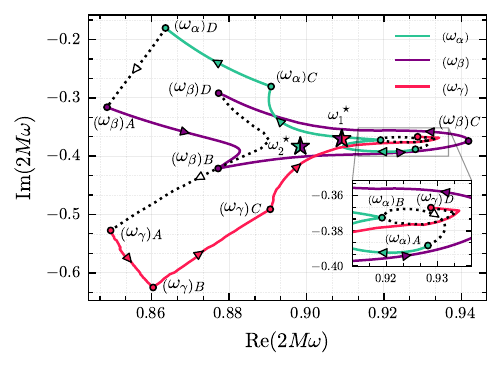}
\caption{Evolution of the QNM frequencies tracked along the closed path $\Gamma_{ABCDA}$, with the last segment shown as dashed lines. 
Here, $\omega_\alpha$ and $\omega_\beta$ denote the QNM frequencies of the $(s,n) = (-2,1)$ and $(-2,2)$ branches, respectively. 
The frequency $\omega_\gamma$ corresponds to the near-horizon branch.
A 3-cyclic non-trivial exchange of states is now exhibited.}
\label{fig:trihysteresis}
\end{figure}

Given the proximity of $\mathrm{EP}_1$ and $\mathrm{EP}_2$, we can construct a closed path in parameter space encircling both EPs at once without leaving their vicinity. Here, the evolution of three different states must be tracked, for each EP induces coalescence of a different pair: $(s,n)=(-2,2)$ with $(-2,1)$ and $(-2,2)$ with near-horizon modes. 
As illustrated in Fig.~\ref{fig:trihysteresis}, this yields the non-trivial exchange of the three states in a cycle: $(-2,1) \to (-2,2) \to \mathrm{NH} \to (-2,1) \to \cdots$.
The resulting cycle of states is, as expected, compatible with the local topology found for a close neighbourhood of $\mathrm{EP}_1$ and $\mathrm{EP}_2$ within $\paramspace \times \{ 2M \omega \}$ (i.e., that of a pair of square-root Riemann sheets).

~

\noindent\textit{Conclusions and discussion.} We have shown that EPs and eigenstate hysteresis arise intrinsically in KN QNM spectrum, illustrated with the study of the $\ell=m=2$ mode. 
The coupling between gravitational and electromagnetic perturbations supplies the second real parameter that an EP requires, so no auxiliary field is needed, in contrast with the Kerr case. 
The hyperboloidal formulation of the coupled Teukolsky system, discretised on a Chebyshev pseudospectral grid, made it possible to resolve the spectrum accurately close to extremality. 

We have located four EPs. Three of them, $\mathrm{EP}_0$, $\mathrm{EP}_1$ and $\mathrm{EP}_2$, are responsible for the instances of the Wigner-Teller effect previously reported~\cite{Dias:2021yju,Dias:2022oqm}. 
The fourth, $\mathrm{EP}_{\rm e}$, comes from a previously unreported repulsion between electromagnetic-led branches (see Fig.~\ref{fig:paths}). 
All four EPs lie in the near-extremal region, from $99.3741\%$ ($\mathrm{EP}_2$) to $99.9740\%$ ($\mathrm{EP}_0$) of extremality, and none is close to the Reissner-Nordstr\"om limit, although $\mathrm{EP}_1$ and $\mathrm{EP}_2$ have a larger relative charge than spin (see Table~\ref{tab:expoints}). 

Quasi-static loops around each EP exchange the coalescing modes (see Figs.~\ref{fig:paths}--\ref{fig:hysteresis} for $\mathrm{EP}_{\rm e}$), and a loop enclosing $\mathrm{EP}_1$ and $\mathrm{EP}_2$ permutes three modes cyclically (see Fig.~\ref{fig:trihysteresis}), as expected from the Riemann-sheet structure around the EPs (see Fig.~\ref{fig:hysteresis}). Together with the square-root closing of the repulsion gaps, this identifies the four EPs as branch-point singularities of the spectrum rather than numerical artefacts, and links BH spectroscopy to the many physical scenarios where EP effects are studied, such as microwave, acoustic and optomechanical systems~\cite{Dembowski:2001prl,PhysRevB.109.144102,10.1093/qjmam/41.3.363}.

Observing hysteresis in a physical process would require closed, adiabatic cycles in the parameter space $\mathcal{P}_{\mathrm{KN}}$. Superradiance changes $(a,Q)$ only monotonically~\cite{Press:1972zz}, so a closed cycle would need, for example, accretion of charged matter, and designing such a process is an open problem. The other $(\ell,m)$ sectors, especially high-order multipoles ($\ell>2$), very likely contain further EPs. An EP close enough to the Kerr limit could leave a signature in the ringdown~\cite{Yang:2025dbn}, and so help constrain small charges in astrophysical BHs, although the EPs found here require charge-to-mass ratios that are unlikely astrophysically.

\noindent\textit{Acknowledgments.}~The authors thank M.G. Benot, V. Cardoso, R. Panosso Macedo, and A. Zenginoğlu for their feedback on different aspects of this work.  M.M.A and A.T.P work was partly funded by the innovation project SOL2024-31517, funded by the EU - Ministerio de Hacienda y Función Pública - Fondos Europeos - Junta de Andalucía – Consejería de Universidad, Investigación e Innovación.
C.F.S. work was supported by contract PID2022-137674NB-I00/AEI/10.13039/501100011033 (Spanish Ministry of Science and Innovation) and 2017-SGR-1469 (AGAUR, Generalitat de Catalunya), and partly supported by the Spanish program Unidad de Excelencia Mar\'ia de Maeztu CEX2020-001058-M, financed by MCIN/AEI/10.13039/501100011033, and by the MaX-CSIC Excellence Award MaX4-SOMMA-ICE. 

~

\noindent \text{\em Data availability.} The data are available from the authors upon reasonable request.


%

\end{document}